\documentclass[12pt]{iopart}
\usepackage{epsfig}
\usepackage{iopams}
\usepackage{epsfig} 
\usepackage{graphicx}
\usepackage{graphics}
\usepackage{subfigure}
\usepackage{verbatim} 
\newcommand{\eq}{\begin{eqnarray}} 
\newcommand{\en}{\end{eqnarray}} 
\newcommand{\ra}{\rangle} 
\newcommand{\la}{\langle}

\newcommand{\bfq}{{\bf q}_{\perp}}
\newcommand{\bfqpr}{{\bf q'}_{\perp}}

\newcommand{\bfk}{{\bf k}_{\perp}}
\newcommand{\bfkpr}{{\bf k}_{\perp}^\prime}

\newcommand{\bfP}{{\bf P}_{\perp}} 
\newcommand{\bfPpr}{{\bf P}_{\perp}^\prime}

\begin{document}
\title{Pion light-front wave function, parton distribution   
and the electromagnetic form factor\\} 

\author{Thomas Gutsche$^1$, 
        Valery E. Lyubovitskij$^{1,2,3}$, 
        Ivan Schmidt$^4$, 
        Alfredo Vega$^{5,6}$
\vspace*{1.2\baselineskip}}
\address{
$^1$ Institut f\"ur~Theoretische Physik, Universit\"at T\"ubingen,
Kepler Center for Astro and Particle Physics, 
Auf der Morgenstelle 14, D--72076 T\"ubingen, Germany\\ 
$^2$ Department of Physics, Tomsk State University,  
634050 Tomsk, Russia\\
$^3$ Mathematical Physics Department, Tomsk Polytechnic University, 
Lenin avenue 30, 634050 Tomsk, Russia\\
$^4$ Departamento de F\'\i sica y Centro Cient\'\i
fico Tecnol\'ogico de Valpara\'\i so (CCTVal), Universidad T\'ecnica
Federico Santa Mar\'\i a, Casilla 110-V, Valpara\'\i so, Chile\\
$^5$ Instituto de F\'isica y Astronom\'ia, 
Universidad de Valpara\'iso,  \\
Avenida Gran Breta\~na 1111, Valpara\'iso, Chile\\
$^6$ Centro de Astrof\'isica de Valpara\'iso, 
Avenida Gran Breta\~na 1111, Valpara\'iso, Chile\\}     
\ead{
thomas.gutsche@uni-tuebingen.de,
valeri.lyubovitskij@uni-tuebingen.de, 
ivan.schmidt@usm.cl, 
alfredo.vega@uv.cl
} 

\date{\today}

\begin{abstract} 

We derive the lowest-order $q\bar q$ valence light-front 
wave function for the pion with $L_z=0$ and $|L_z|=1$, 
which reproduces its valence parton distribution and 
the electromagnetic form factor consistent with data. 

\end{abstract}

\vskip 1cm



\maketitle

\section{Introduction}

The main purpose of this paper is to derive 
the lowest-order $q\bar q$ valence light-front wave function (LFWF) 
for the pion with $L_z=0$ and $|L_z|=1$, 
which can generate a pion parton distribution function (PDF) 
$q_\pi(x)$ and the electromagnetic form factor $F_\pi(Q^2)$ 
consistent with data. 
The present study is based on data on the pion 
electromagnetic form factor~\cite{Volmer}-\cite{Tadevosyan:2007yd} 
and the model-independent result for the pion PDF at the initial scale 
$\mu_0 = 0.63$ GeV 
\eq 
x q_\pi(x,\mu_0) = N_\pi x^\alpha \, (1-x)^\beta \, (1+\gamma x^\delta) 
\en 
derived in Ref.~\cite{Aicher:2010cb}. The result of~\cite{Aicher:2010cb} 
rests on an updated analysis of the E615 data on the cross section of the 
Drell-Yan (DY) process $\pi^- N \to \mu^+ \mu^- X$~\cite{Conway:1989fs}  
including next-to-leading logarithmic (NLL) threshold resummation effects. 
The coefficients $\alpha$, $\beta$, $\gamma$ and $\delta$ are free 
parameters, which for the best fit of Ref.~\cite{Aicher:2010cb} 
are reported as 
\eq\label{parameters} 
\alpha = 0.70 \,, \ 
\beta  = 2.03\,, \ 
\gamma = 13.8\,, \ 
\delta = 2 \,. 
\en   
The normalization constant $N_\pi$ is fixed by the condition 
\eq 
\int\limits_0^1 dx q_\pi(x,\mu_0) = 1 \,. 
\en 
These parameters were fixed by evolving the pion PDF to the higher 
scale $\mu = 4$ GeV, where the E615 data were extracted and analyzed 
leading to 
\eq 
q_\pi(x,\mu = 4 \ {\rm GeV}) \sim (1-x)^{2.34}
\en 
for large $x$. 
The main result of  Ref.~\cite{Aicher:2010cb} is that 
the pion PDF at the initial scale and for large $x$ with 
\eq 
q_\pi(x,\mu_0) \sim (1-x)^{2.03} 
\en 
is considerably softer when compared to the next-to-leading order 
analysis of the DY process performed previously (see e.g. 
Refs.~\cite{Sutton:1991ay}-\cite{Wijesooriya:2005ir}). Note that 
the result of Ref.~\cite{Aicher:2010cb} is in agreement with 
calculations based on the Dyson-Schwinger approach~\cite{Hecht:2000xa}.  
The pion PDF has also been subject of detailed analyses in quark models, 
as for example the one performed 
in the chiral quark model~\cite{Broniowski:2007si}. 
As in the series of NLO analysis that have been made without the inclusion of 
soft-gluon resummation, the analysis of Ref.~\cite{Broniowski:2007si} was 
based on a harder PDF at the scale $\mu = 4$ GeV with 
\eq 
q_\pi(x,\mu = 4 \ {\rm GeV}) \sim (1-x) \,.
\en 
Using the prediction of the chiral quark model 
for the pion PDF at the initial scale with $q_\pi(x,\mu_0) = 1$, 
this initial scale $\mu_0 \simeq 313$ MeV was determined 
by evolving the $q_\pi(x,\mu)$ down from $\mu = 4$ GeV. 
Recently~\cite{Gutsche:2013zia} we could confirm the results 
of Ref.~\cite{Broniowski:2007si}. 
In particular, we applied a soft-wall AdS/QCD 
model~\cite{Brodsky:2007hb}-\cite{Vega:2010ns} 
for the analysis of the evolution of the 
pion PDF. As in Ref.~\cite{Broniowski:2007si}, 
the pion PDF obtained in soft-wall 
AdS/QCD should be considered at the initial scale. At $\mu = 4$ GeV the result 
of~\cite{Gutsche:2013zia} coincides with the result of the chiral quark 
model~\cite{Broniowski:2007si} and is very well approximated by the form 
\eq 
q_\pi(x,\mu = 4 \ {\rm GeV}) = 1.108 \, x^{-0.381} \, (1-x)^{1.323} \,. 
\en 
A nice feature of the soft-wall AdS/QCD approach
is that it allows for an extraction of the hadronic LFWF. 
The idea to extract LFWFs from AdS/QCD was 
originally suggested in Ref.~\cite{Brodsky:2007hb} considering 
the pion electromagnetic form factor in two approaches -- AdS/QCD and 
light-front QCD. In a series of papers~\cite{Brodsky:2007hb,%
Vega:2009zb,Branz:2010ub,Brodsky:2011xx,Gutsche:2012ez}   
this problem was further discussed in detail. 
The pionic $L_z=S_z=0$ LFWF extracted from AdS/QCD reads~\cite{Brodsky:2011xx}
\eq 
\psi^{\rm AdS}_\pi(x,\bfk) = \frac{4\pi}{\kappa}
 \frac{\sqrt{\log(1/x)}}{1-x} 
\exp\biggl[- \frac{\bfk^2}{2\kappa^2} 
\frac{\log(1/x)}{(1-x)^2} \biggr] \,, 
\en
where $\kappa = 350$ MeV is the dilaton scale 
parameter~\cite{Gutsche:2013zia}. 
Note that the derived LFWF is not symmetric under the exchange 
$x \to 1-x$, since it results from the matching of the matrix 
element for the bare electromagnetic current between the dressed 
LFWF in LF QCD and the dressed electromagnetic current between 
hadronic wave functions in AdS/QCD. The recent analysis of 
Ref.~\cite{Aicher:2010cb} shows, however, that inclusion of the NLL  
threshold resummation effects in the $\pi^- N \to \mu^+ \mu^- X$ 
cross section leads to a considerably softer pion PDF at the initial 
(low) scale. This obviously calls for a reconsideration of the pionic LFWF.

\section{Light-front wave function of the pion and applications} 

In this section we offer an updated pion LFWF which, first, produces 
the PDF extracted in the analysis of~\cite{Aicher:2010cb} and, second, 
results in a pion electromagnetic form factor consistent with current 
data. Following Ref.~\cite{Burkardt:2002uc,Ji:2003yj,Pasquini:2014ppa} 
we construct the lowest quark-antiquark Fock component LFWF for the pion, 
which includes the superposition of three components with 
the total quark orbital angular momentum $L_z = 0, \pm 1$. Considering for 
example a positively charged pion with momentum 
$P = (P^+, (M_\pi^2+\bfP^2)/P^+, \bfP)$ the light-front pion state 
reads~\cite{Burkardt:2002uc,Ji:2003yj,Pasquini:2014ppa} 
\eq 
\hspace*{-2.5cm}
|\pi^+(P^+,\bfP)\ra &=& \ |\pi^+(P^+,\bfP)\ra_{L_z=0} + 
|\pi^+(P^+,\bfP)\ra_{|L_z|=1} \,,\nonumber\\
\hspace*{-2.5cm}
|\pi^+(P^+,\bfP)\ra_{L_z=0}  &=&  
\int \frac{d^2\bfk}{16\pi^3} \, \frac{dx}{\sqrt{x (1-x)}} \, 
\psi_\pi^{(0)}(x,\bfk) \, |x, \bfk; P^+, \bfP, L_z=0\ra\,, \\ 
\hspace*{-2.5cm}
|\pi^+(P^+,\bfP)\ra_{|L_z|=1}   &=& 
\int \frac{d^2\bfk}{16\pi^3} \, \frac{dx}{\sqrt{x (1-x)}} \, 
\psi_\pi^{(1)}(x,\bfk) \nonumber\\ 
&\times& \frac{1}{\sqrt{2}} \, \Big(k_\perp^+\, 
|x, \bfk; P^+, \bfP, L_z=1\ra 
+ k_\perp^-\, |x, \bfk; P^+, \bfP, L_z=-1\ra \Big)\,,  \nonumber 
\en 
where $k_\perp^\pm = k_1 \pm ik_2$, 
$\psi_\pi^{(0)}(x,\bfk)$ and 
$\psi_\pi^{(1)}(x,\bfk)$ are the pion $L_z=0$ and $|L_z|=1$ 
phenomenological LFWFs, respectively;  
$|x, \bfk; P^+, \bfP, L_z\ra)$ is the $u\bar d$ Fock state with total 
quark orbital angular momentum $L_z$. The Fock states 
$|\pi^+(P^+,\bfP)\ra_{L_z}$, $|x, \bfk; P^+, \bfP, L_z\ra$ 
and LFWFs $\psi_\pi^{(i)}(x,\bfk)$
are normalized according to 
\eq 
\hspace*{-2.5cm}
\la x', \bfkpr; P^{+'}, \bfPpr, L_z' | x, \bfk; P^+, \bfP, L_z\ra 
&=& 2P^+ 2 x (1-x) \, (2\pi^3)^2 \, \delta(P^+-P^{+'}) \, 
\delta(x-x') 
\nonumber\\
\hspace*{-2.5cm}
&\times&\delta^2(\bfk-\bfkpr) \, \delta^2(\bfP-\bfPpr) 
\, \delta_{L_zL_z'} 
\,, \nonumber\\
\hspace*{-2.5cm}
\la\pi^+(P^{+'},\bfPpr)|\pi^+(P^+,\bfP)\ra &=& 2P^+ \, (2\pi)^3 
\, \delta(P^+ - P^{+'}) \, \delta^2(\bfP-\bfPpr) 
\en 
and 
\eq\label{norm_cond}
1 &=& \int\limits_0^1 dx q_\pi(x)\,, \nonumber\\
q_\pi(x) &=&\int\frac{d^2\bfk}{16\pi^3} \, 
\biggl[|\psi_\pi^{(0)}(x,\bfk)|^2 \,+\, \bfk^2 
|\psi_\pi^{(1)}(x,\bfk)|^2 \biggr] \,. 
\en 
In last equation $q_\pi(x)$ is the pion valence parton distribution 
function (PDF). A representation of the $u\bar d$ Fock states 
$| x, \bfk; P^+, \bfP, L_z\ra$ in terms of creation operators 
of $u$- and $\bar d$-quarks is given in Appendix~A.   
We restrict to the isospin limit, therefore 
$q_\pi(x)$ coincides with the distribution of $u$ and $\bar d$- 
quarks in the pion 
\eq 
q_\pi(x) = u_v^\pi(x) = \bar d_v^\pi(x) \,. 
\en 
The pion electromagnetic form factor $F_\pi(Q^2)$ 
(see e.g. details in Ref.~\cite{Brodsky:2007hb}) 
is defined as 
\eq\label{Fpi_em} 
\la \pi^+(P^+,\bfq)|J^+(0)|\pi^+(P^+,{\bf 0}_\perp)\ra = 
2P^+ F_\pi(Q^2) 
\en  
where we choose the light-front frame coordinates of 
the initial pion $(P)$ and photon $(q)$ as~\cite{Brodsky:2007hb}  
\eq 
P = (P^+, M_\pi^2/P^+,{\bf 0}_\perp) \,, \quad 
q = (0, 2qP/P^+, \bfq)  \,. 
\en  
Here $J^+(0) = \sum\limits_{q} e_q \, \bar q(0) \gamma^+ q(0)$ 
is the electromagnetic current operator (see its representation 
in terms of creation and annihilation operators of quarks and 
antiquarks in Ref.~\cite{Brodsky:2007hb} and also in Appendix~A).  
Evaluating Eq.~(\ref{Fpi_em}) gives the representation of the 
pion electromagnetic form factor in terms of the LFWFs 
$\psi_\pi^{(0)}(x,\bfk)$ and 
$\psi_\pi^{(1)}(x,\bfk)$ (Drell-Yan-West formula~\cite{Drell:1969km}) 
\eq 
\hspace*{-2cm}
F_\pi(Q^2)  &=& 
\int\limits_0^1 dx \, {\cal H}_\pi(x,Q^2) \,, \nonumber\\ 
\hspace*{-2cm}
{\cal H}_\pi(x,Q^2) &=& 
\int\frac{d^2\bfk}{16\pi^3} \, 
\biggl[ \psi^{(0)\dagger}_\pi(x,\bfk') \, 
     \psi^{(0)}_\pi(x,\bfk) \,+\, 
     \bfk' \bfk \, 
     \psi^{(1)\dagger}_\pi(x,\bfk') \, 
     \psi^{(1)}_\pi(x,\bfk) 
\biggr]
\en 
where 
$\bfk' = \bfk + (1-x) \bfq$ and $Q^2 = \bfq^2$ and 
\eq 
{\cal H}_\pi(x,Q^2) 
= H_\pi(x,\xi=0,Q^2) - H_\pi(-x,\xi=0,Q^2) 
\en 
is the nonforward parton density (NPD)~\cite{Radyushkin:1998rt} 
for the pion evaluated at zero skewness $\xi = 0$. At $Q^2=0$ 
the relation ${\cal H}_\pi(x,0) = q_\pi(x)$ follows.  
We choose the LFWFs $\psi_\pi^{(0)}(x,\bfk)$ and 
$\psi_\pi^{(1)}(x,\bfk)$ as  
\eq 
\hspace*{-2cm}
\psi_\pi^{(0)}(x,\bfk) &=& \frac{4\pi N_0}{\kappa} \, 
\frac{\sqrt{\log(1/x)}}{1-x} \, 
\sqrt{f(x) \, \bar f(x)} \, \exp\biggl[- \frac{\bfk^2}{2\kappa^2} 
\frac{\log(1/x)}{(1-x)^2} \, \bar f(x) \biggr] \,, \nonumber\\
\hspace*{-2cm}
\psi_\pi^{(1)}(x,\bfk) &=& \frac{4\pi N_1}{\kappa^2} \, 
\frac{\sqrt{\log^3(1/x)}}{(1-x)^2} \, 
\sqrt{f(x) \, \bar f^3(x)} \, \exp\biggl[- \frac{\bfk^2}{2\kappa^2} 
\frac{\log(1/x)}{(1-x)^2} \, \bar f(x) \biggr] 
\en 
to guarantee the correct scaling behavior of the pion PDF, NPD 
and form factor due to the contribution of 
$\psi^{(0)}_\pi(x,\bfk)$ and $\psi^{(1)}_\pi(x,\bfk)$. 
In particular, $\psi^{(0)}_\pi(x,\bfk)$ 
produces the following scalings of 
$q_\pi(x) \sim (1-x)^2$ and ${\cal H}_\pi(x,Q^2) \sim (1-x)^2$ 
for large $x \to 1$ and $F_\pi(Q^2) \sim 1/Q^2$ for large $Q^2 \to \infty$.
The LFWF $\psi^{(1)}_\pi(x,\bfk)$ results in 
$q_\pi(x) \sim (1-x)^5$, ${\cal H}_\pi(x,Q^2) \sim (1-x)^5$ 
for $x \to 1$ and $F_\pi(Q^2) \sim 1/Q^4$ for $Q^2 \to \infty$. 
The functions $f(x)$ and $\bar f(x)$ are specified as 
\eq 
f(x) = x^{\alpha-1} \, (1-x)^\beta \, 
(1+\gamma x^\delta) \,, \quad\quad 
\bar f(x) = x^{\bar\alpha} \, (1-x)^\beta \, 
(1+\bar\gamma x^{\bar\delta}) \,.
\en 
The $\alpha$, $\beta$, $\gamma$, $\delta$, 
      $\bar\alpha$, $\bar\gamma$ and $\bar\delta$ 
are free parameters and $N_0$ and $N_1$ are the normalization factors.  

A straightforward calculation gives 
\eq\label{pionGPD} 
{\cal H}_\pi(x,Q^2) &=& N_0^2 f(x) 
\exp\biggl[- \frac{Q^2}{4\kappa^2} \, 
\log(1/x) \, \bar f(x) \biggr] \nonumber\\
&\times&\biggl[ 
1 + \biggl(\frac{N_1}{N_0}\biggr)^2 
\bar f(x) \log(1/x) 
\biggl( 1- \frac{Q^2}{4\kappa^2} \log(1/x) \, \bar f(x) \biggr) \biggr] \,. 
\en 
The parameter $N_0$ is fixed from the normalization 
condition~(\ref{norm_cond}), while the ratio $N_1/N_0$ is a free parameter. 

The function $f(x)$ is constrained by the pion PDF fixed through 
Ref.~\cite{Aicher:2010cb}
\eq 
q_\pi(x) \sim f(x)\,,  
\en
while the function $\bar f(x)$ is fixed from the analysis of 
the pion electromagnetic form factor (as will be shown below). 
Note that the improved LFWF $\psi_\pi(x,\bfk)$ reduces
to the AdS/QCD LFWF $\psi^{\rm AdS}_\pi(x,\bfk)$ 
in the limit $f(x) = \bar f(x) = 1$. 
By construction the functions $f(x)$ and $\bar f(x)$ have the same 
scaling behavior of $(1-x)^\beta$ at large $x$ to guarantee the 
correct $1/Q^2$ power scaling of the pion form factor at large $Q^2$ 
independent on the value of $\beta$ (see below). 

\begin{figure}
\begin{center}

\vspace*{.5cm}
\epsfig{figure=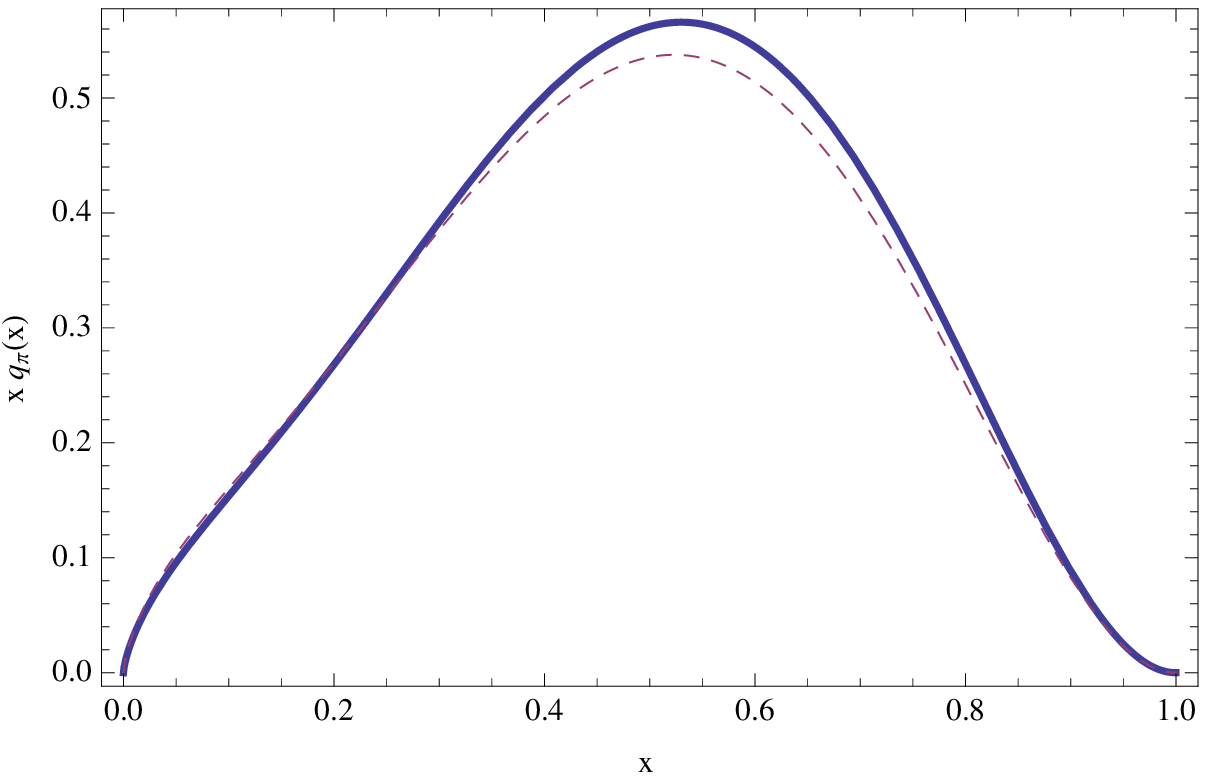,scale=.75}

\noindent
\caption{Pion valence PDF $x q_\pi(x)$. The thick line 
corresponds to the leading $L_z=0$ LFWF contribution, 
while the dashed line includes both $L_z=0$ and 
$|L_z|=1$ LFWF contributions. 
\label{fig:pion_pdf}}

\vspace*{2cm}
\epsfig{figure=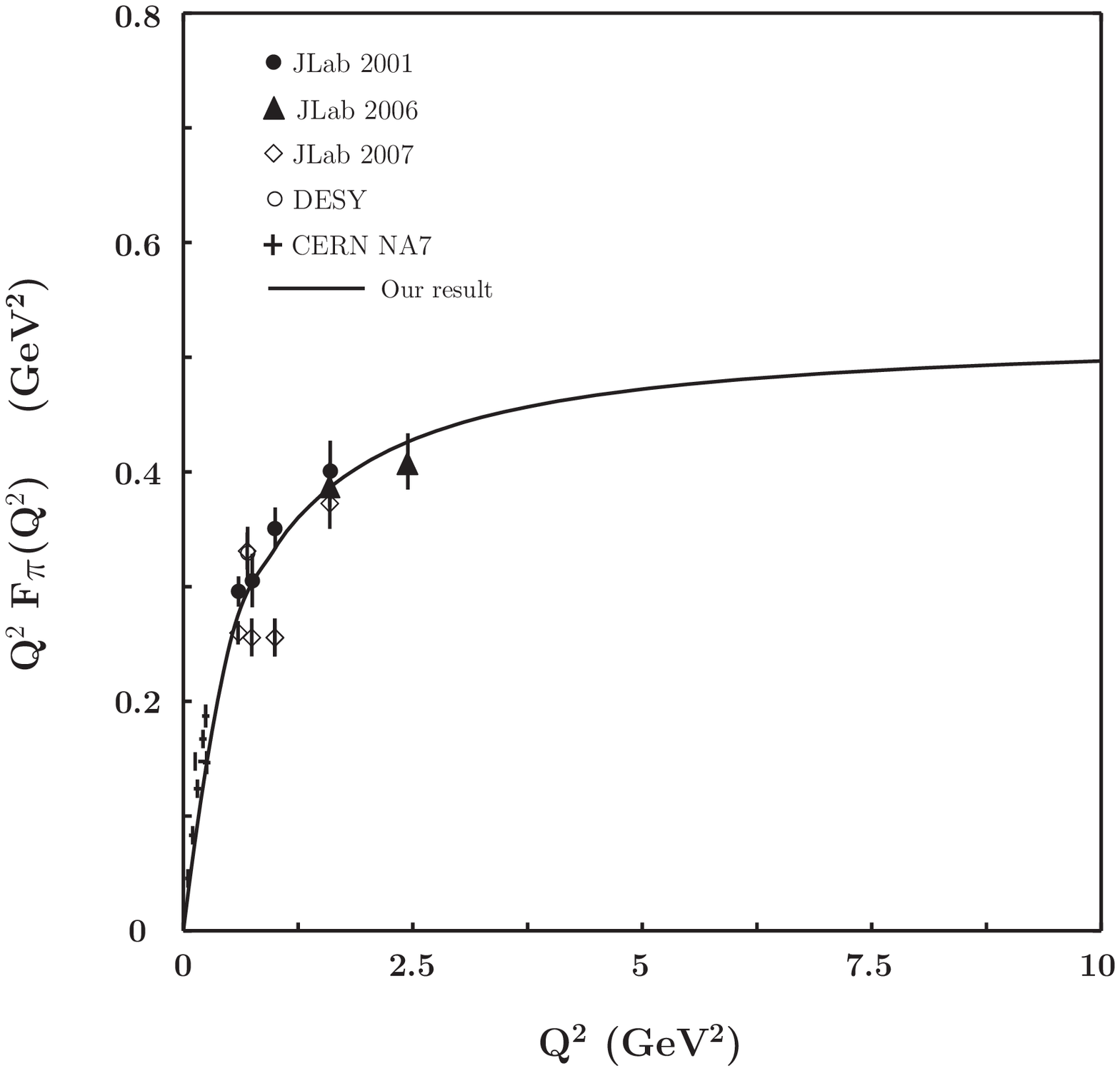,scale=.5}
\vspace*{-2cm}
\noindent
\caption{Pion form factor $Q^2 F_\pi(Q^2)$. Data taken 
from~\cite{Volmer}-\cite{Tadevosyan:2007yd}.  
\label{fig:pion_ff}}
\end{center}
\end{figure}

\begin{figure}
\begin{center}
\epsfig{figure=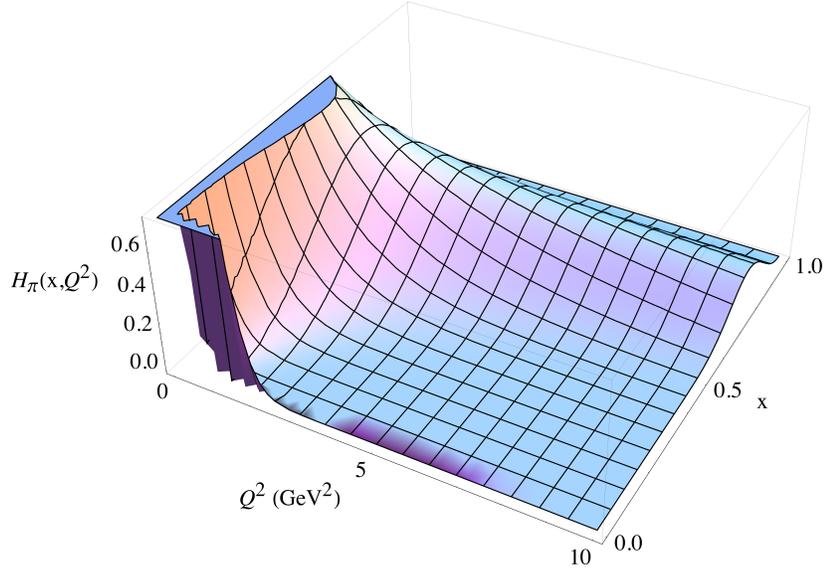,scale=.85}
\noindent
\end{center}
\vspace*{-.5cm}
\caption{Pion nonforward parton density ${\cal H}_\pi(x,Q^2)$ 
for $x = [0, 1]$ \\ and $Q^2$ = [0, 10 GeV$^2$]. 
\label{fig:pion_GPD}}
\end{figure}

Next we discuss the choice of the free parameters 
$\alpha$, $\beta$, $\gamma$, $\bar\alpha$ and $\bar\gamma$. 
The values for $\alpha$, $\beta$, $\gamma$ (central values) 
are taken from Ref.~\cite{Aicher:2010cb} (see Eq.~(\ref{parameters})).  
The parameters $\bar\alpha$, $\bar\gamma$ and $\bar\delta$ are 
fixed by a fit to the pion electromagnetic form factor resulting in:  
\eq
\bar\alpha = 0.15\,, \ 
\bar\gamma = 2.5\,,  \ 
\bar\delta = 2\,. 
\en 
Also we use $N_1/N_0 = 0.3$. 
As stressed before, the pion form factor has the correct $1/Q^2$ scaling 
for large $Q^2$, consistent with quark counting rules. It is also 
independent of $\beta$, which governs the power scaling $(1-x)^\beta$ 
of the pion PDF and GPD at large~$x$. Here we use $\beta = 2.03$  
extracted from the updated analysis of the E615 experimental 
data~\cite{Conway:1989fs} including NLL threshold resummation 
effects~\cite{Aicher:2010cb}. The choice of 
our parameters is constrained by data on 
the electromagnetic radius of the pion with 
$r_\pi^{\rm exp}= 0.672 \pm 0.008$ fm~\cite{Agashe:2014kda}. 
In particular, we are able to reproduce the 
central value of $r_\pi = 0.672$ fm. 
In Fig.~\ref{fig:pion_pdf} we show our result for the pion valence PDF, 
indicating the leading contribution due to $L_z=0$ LFWF (thick line), which 
fits the result of Ref.~\cite{Aicher:2010cb}, and the
total result including both $L_z=0$ and $|L_z|=1$ LWFs (dashed line). 
The total result can reproduce the result of 
Ref.~\cite{Aicher:2010cb} for the pion PDF, 
when two of our parameters are slightly changed 
($\alpha = 0.70 \to 0.75$ and $\bar\delta = 2 \to 1.65$).    
In Fig.~\ref{fig:pion_ff} we show our result for the pion 
electromagnetic form factor multiplied by $Q^2$ and compare it to data. 
In Fig.~\ref{fig:pion_GPD} we show a three-dimensional plot of 
the pion NPD as function of $x$ in the interval from 0 to 1 and $Q^2$ from 
0 to 10 GeV$^2$. 

In conclusion, we state again the main result of this report. 
We first take into account the result of Ref.~\cite{Aicher:2010cb}, 
which produces a considerably softer pion PDF when the NLL threshold 
resummation effects in the cross section of $\pi^- N \to \mu^+ \mu^- X$  
process are included.  We derive an improved pion LFWF at this order of 
accuracy. This LFWF reproduces several fundamental 
properties of the pion consistent with data --- 
valence parton distribution, electromagnetic form factor 
and radius. The derived LFWF is parametrized by two functions 
$f(x)$ and $\bar f(x)$ depending on the light-cone variable $x$, which 
have a clear physical interpretation. In particular, the function $f(x)$ 
is related to the pionic PDF at the initial scale derived in 
Ref.~\cite{Aicher:2010cb}. The function $\bar f(x)$ defines the $Q^2$ 
dependence of the pion GPD and the electromagnetic form factor. 
Both functions $f(x)$ and $\bar f(x)$ have exactly the same power 
scaling $(1-x)^\beta$ at large $x \to 1$. This behavior provides 
the correct power scaling of the pion form factor. In particular, 
the $L_z=0$ LFWF $\psi^{(0)}(x,\bfk)$ and 
$|L_z|=1$ LFWF $\psi^{(1)}(x,\bfk)$ give the following scaling of the 
pion form factor at large $Q^2 \to \infty$ 
\eq 
F_\pi^{(0)}(Q^2) &\sim& \int\limits_0^1 dx (1-x)^\beta 
\exp\biggl[- \frac{Q^2}{4\kappa^2} \, (1-x)^{\beta + 1} \biggr] 
\sim \frac{1}{Q^2} \,, \nonumber\\
F_\pi^{(1)}(Q^2) &\sim& \int\limits_0^1 dx (1-x)^{2\beta+1}  
\exp\biggl[- \frac{Q^2}{4\kappa^2} \, (1-x)^{\beta + 1} \biggr] 
\sim \frac{1}{Q^4} \,, \nonumber\\
\en  
respectively, which are consistent with quark counting rules. 
Using data on the pion electromagnetic form factor we constrain 
the form of the function $\bar f(x)$ and also produce the result 
for the pion nonforward parton density ${\cal H}_\pi(x,Q^2)$ at 
zero skewness. 

\section*{Acknowledgments}

The authors thank Werner Vogelsang for useful discussions. 
This work was supported by the DFG under Contract No. LY 114/2-1, 
by the German Bundesministerium f\"ur Bildung und Forschung (BMBF) 
under Grant No. 05P12VTCTG, by Marie Curie Reintegration Grant IRG 256574, 
by CONICYT (Chile) Research Project No. 80140097,  
by CONICYT (Chile) under Grant No. 7912010025, 
by FONDECYT (Chile) under Grant No. 1140390 and No. 1141280,    
by Tomsk State University Competitiveness Improvement Program and the 
Russian Federation program ``Nauka'' (Contract No. 0.1526.2015, 3854).  
V.E.L. would like to thank Departamento de F\'\i sica y Centro
Cient\'\i fico Tecnol\'ogico de Valpara\'\i so (CCTVal), Universidad
T\'ecnica Federico Santa Mar\'\i a, Valpara\'\i so, Chile and 
Instituto de F\'isica y Astronom\'ia, Universidad de Valpara\'iso, Chile 
for warm hospitality.

\newpage 

\appendix 
\section{Quark-antiquark Fock states and electromagnetic 
transition operator in light-front frame} 

Here we specify quark-antiquark Fock states and 
electromagnetic transition operator in terms of creation and annihilation 
operators in the light-front formalism. The $u\bar d$ Fock states 
with specific value of $L_z=0, \pm 1$ are  
\eq
|x, \bfk; P^+, \bfP, L_z=0\ra &=&
\frac{1}{\sqrt{6}} \, \sum_{a=1}^{3} \,
\Big(b^{\dagger a}_{u\uparrow}(1) d^{\dagger a}_{d\downarrow}(2) 
-    b^{\dagger a}_{u\downarrow}(1) 
     d^{\dagger a}_{d\uparrow}(2)\Big) \, |0\ra
\,,\nonumber\\
|x, \bfk; P^+, \bfP, L_z=-1\ra &=&
\frac{1}{\sqrt{3}} \, \sum_{a=1}^{3} \,
     b^{\dagger a}_{u\uparrow}(1) 
     d^{\dagger a}_{d\uparrow}(2) \, |0\ra\,, \nonumber\\
|x, \bfk; P^+, \bfP, L_z=1\ra &=&
\frac{1}{\sqrt{3}} \, \sum_{a=1}^{3} \,
    b^{\dagger a}_{u\downarrow}(1) 
     d^{\dagger a}_{d\downarrow}(2)  \, |0\ra\,, 
\en 
where $1 \doteq xP^+,\bfk+x\bfP$ and $2 \doteq (1-x)P^+,-\bfk+(1-x)\bfP$. 
Here $b^{\dagger a}_\lambda(b^a_\lambda)$ and
$d^{\dagger a}_\lambda(d^a_\lambda)$ are the
creation (annihilation) operators of $u$-quark and $\bar d$ quark
with color $a$, respectively,
obeying the nonvanishing anticommutation relations 
\eq
& &
\{b^a_\lambda(k^+,\bfk),b^{\dagger a'}_{\lambda'}(k^{+'},\bfkpr)\} = 
\{d^a_\lambda(k^+,\bfk),d^{\dagger a'}_{\lambda'}(k^{+'},\bfkpr)\}\nonumber\\
&=& 
2 k^+ \, \delta_{\lambda\lambda'} \, \delta^{aa'} \,
\delta(k^+ - k^{+'}) \, \delta^2(\bfk-\bfkpr)\,. 
\en
The electromagnetic current transition operator in terms of 
creation and annihilation operators of $u$ and $\bar d$ quarks 
read as~\cite{Brodsky:2007hb} 
\eq 
\hspace*{-1.5cm}
J^+(0) &=& \int \frac{dq^+ d^2\bfq}{(2\pi)^3 \sqrt{2q^+}} 
\int \frac{dq^{'+} d^2\bfqpr}{(2\pi)^3 \sqrt{2q^{'+}}} 
\, \sum\limits_{a=1}^3
\biggl[ 
e_u b^{\dagger a}_{u\uparrow}(q) 
    b^{a}_{u\uparrow}(q') + 
e_u b^{\dagger a}_{u\downarrow}(q) 
    b^{a}_{u\downarrow}(q') \nonumber\\
\hspace*{-1.5cm}
&+& 
e_{\bar d} d^{\dagger a}_{d\uparrow}(q) 
           d^{a}_{d\uparrow}(q') +  
e_{\bar d} d^{\dagger a}_{d\downarrow}(q) 
           d^{a}_{d\downarrow}(q')  
\biggr]
\en

\end{document}